\documentclass[a4paper,aps,prd,showpacs,nofootinbib,preprintnumbers,amsmath,amssymb,mcite,superscriptaddress,12pt]{revtex4-1}
\usepackage{epsfig}
\usepackage{subfig}
\usepackage[table]{xcolor}
\usepackage{graphicx}
\usepackage{dcolumn}
\usepackage{bm}
\usepackage{esvect}
\usepackage{tikz}
\usepackage{accents}
\usepackage{ulem}

\usepackage{hyperref}
\hypersetup{colorlinks=true,linkcolor=magenta,anchorcolor=green,citecolor=cyan,filecolor=black,menucolor=black,urlcolor=brown}
\usepackage{slashed}

\newcommand{\be}{\begin{equation}}
\newcommand{\ee}{\end{equation}}
\newcommand{\ba}{\begin{eqnarray}}
\newcommand{\ea}{\end{eqnarray}}

\begin{document}

\preprint{IPHT-t18/004}

\title{Screening the Higgs portal}

\author{Philippe Brax}
\affiliation{Institut de Physique Th\'eorique, Universit\'e  Paris-Saclay, CEA, CNRS, F-91191 Gif-sur-Yvette Cedex, France}
\author{Clare Burrage}
\affiliation{School of Physics and Astronomy, University of Nottingham,
Nottingham, NG7 2RD, UK}
\begin{abstract}
Light scalar fields that couple to matter through the Higgs portal mediate long range fifth forces. We show how the mixing of a light scalar with the Higgs field can lead to this fifth force being screened around  macroscopic objects.   This behaviour can only be seen by considering both scalar fields as dynamical, and is missed if the mixing between the Higgs field and the scalar field is not taken into account.    We explain under which conditions the naive ``integrating out" procedure  fails, i.e. when the mass matrix of the Higgs-scalars system has a nearly vanishing mass eigenvalue. The resulting flat direction in field space can be lifted at the quadratic order in the presence of matter and the resulting fifth force mediated by the Higgs portal can be screened either when the gravitating objects are large enough or their surface Newton potential exceeds a threshold.  Finally  we discuss the implications of these results for nearly massless relaxion models.
\end{abstract}

\maketitle
\section{Introduction}
One highly-developed  approach to searching for new physics, which is agnostic about the nature of the full underlying theory, is to look for new fields that couple to the Standard Model through `portal operators' \cite{Beacham:2019nyx}. The most minimal, and therefore arguably most well motivated of these introduce no new scales into the theory, and therefore the resulting interactions are not necessarily suppressed by a high scale.  In this work we will focus specifically on the introduction of an additional light scalar field, which couples to the Standard Model through the `Higgs portal' \cite{Patt:2006fw,OConnell:2006rsp} via a standard Yukawa coupling. New scalar fields of this type  are motivated by a wide range of fundamental open questions, including the nature of dark matter \cite{Silveira:1985rk,McDonald:1993ex,Burgess:2000yq,Pospelov:2007mp,Matos:2008ag}, the nature of dark energy \cite{Davoudiasl:2004be,Bertolami:2007wb,Dimopoulos:2018eam} and the hierarchy problem \cite{Englert:2013gz,Graham:2015cka}, and  can also be part of a mechanism of electroweak baryogenesis \cite{Kuzmin:1985mm,Curtin:2014jma,Kozaczuk:2019pet}.

In this context, the coupling of the scalar field to matter {is} simply induced by the `Higgs portal' coupling of the scalar field to the Higgs field, together with the Yukawa  coupling of the Higgs field to matter
\be
{\cal L}\supset \lambda_\psi h \bar \psi \psi
\ee
where we simplify the description by considering matter as a Dirac field $\psi$ with a Dirac mass term $m_\psi= \lambda_\psi \langle h\rangle $ when the Higgs field acquires a non-vanishing vacuum expectation value (vev) $\langle h\rangle $. This vev triggers the electroweak symmetry breaking and results from
the low energy Higgs potential, i.e.  a polynomial  with self-interactions up to order four only,  whose coefficients depend on  the light scalars.
The Higgs potential reads explicitly
\be
V(h)= -\frac{\mu^2}{2} h^2 + \frac{\lambda}{4}h^4
\ee
where the $\mu$ term, which drives the electroweak symmetry breaking, is now field dependent, and thus scalar fields coupled through the Higgs portal can help to explain the origin of the electroweak scale \cite{Englert:2013gz,Steele:2013fka}. This mechanism is at the heart of ``relaxion" models \cite{Graham:2015cka} where the rolling of the scalars down their interaction potential  eventually triggers the Higgs symmetry breaking. Experimental constraints on the relaxion model are discussed in Ref. \cite{Flacke:2016szy}.
Bounds on scalar fields coupled through the Higgs portal arise from many different measurements including particle colliders and  fixed target experiments. A review of these experimental searches and their future prospects can be found in \cite{Lanfranchi:2020crw}. Constraints also arise from astrophysical observations including the abundance of light elements produced by Big Bang Nucleosynthesis \cite{Fradette:2017sdd} and the energy loss of supernovae \cite{Krnjaic:2015mbs}.

If these scalars couple to matter through the Higgs portal only,  one phenomenological consequence is  the existence of   fifth forces. If the scalars are light these fifth forces are subject to constraints from a wide range of experimental searches \cite{Adelberger:2003zx,Adelberger:2009zz,Piazza:2010ye}.
In this work we will show how particular choices of portal couplings and potential for the light scalars can enable them to hide from experimental searches for fifth forces. This follows from  the  non-linear regime of the theory in the presence of matter.  The mechanism through which the fifth forces are suppressed is therefore similar in nature to the screening mechanisms that have been widely studied for theories of dark energy and modified gravity \cite{Dehnen:1992rr,Gessner:1992flm,Damour:1994zq,Khoury:2003aq,Khoury:2003rn,Gubser:2004uf,Pietroni:2005pv,Olive:2007aj,Nicolis:2008in,Babichev:2009ee,Hinterbichler:2010es,Hinterbichler:2011ca,Joyce:2014kja}, for a review of the constraints on theories with screening see Refs. \cite{Burrage:2017qrf,Ishak:2018his,Noller:2020afd}.  However the mechanism we present here is a novel way of suppressing a scalar mediated fifth force arising from the mixing with the Higgs field.\footnote{Scalar dark energy models, and scalar tensor theories of modified gravity typically couple to matter through a conformal rescaling of the metric.  In Ref. \cite{Brax:2014baa,Burrage:2018dvt} it was shown that, at leading order, these theories are equivalent to scalar fields that couple to matter through the Higgs portal.}

We will uncover three situations. The first is when the mass eigenstates of the Higgs-scalars system in vacuum  are all  positive, the presence of matter in the environment, e.g. a macroscopic matter distribution as appears in laboratory experiments, leads to a linear response theory and the coupling between the scalars and matter depends on the mixing angle between the Higgs field and the scalars. In this case, the vevs of the Higgs and scalar fields are linearly shifted by the matter density inside a massive body and the interaction mediated by the Higgs field  between  massive objects is proportional to their masses resulting in  Yukawa type fifth forces of the standard form. The second case occurs when   one of the eigenstates of the Higgs-scalar system becomes massless. In this case the Higgs-mediated interaction becomes of infinite range in vacuum. We find  then that the linear response theory breaks down. Non-linear effects start dominating as the effective potential  can be parameterised as
\be
U(\delta \phi)\simeq \frac{1}{6} U^{\prime\prime\prime} (\delta \phi)^3 - \frac{\beta_\phi}{m_{\rm Pl}} \delta \phi\rho
\ee
where $U^{\prime\prime\prime}$ is the third derivative of the scalar potential along the massless direction defined by $\delta\phi$ and $\beta_\phi$ is the effective coupling of the scalar to matter of density $\rho$.
Stabilisation of the scalar can only happen when $U^{\prime\prime\prime}$ and $\beta_\phi$ have the same sign. When this is not the case, the presence of matter destabilises the vacuum, we discuss this possibility further in Appendix \ref{app:D}.

We find that the non-linear stabilisation of the scalar field parameterising the flat direction, i.e. the massless eigen-direction in field space, induces the screening of the corresponding long range force. Screening can happen three different ways. The first one and the most common is when the scalar field acquires a large mass at the new minimum of the effective potential. When this happens, the fifth force is Yukawa suppressed and gravitational tests are easily satisfied. The second and third ways are new in the context of the Higgs portal. Indeed when either the massive bodies are large enough or their surface Newton potential is also large enough, the effects of the fifth force are reduced.
These situations could be of interest to cosmology where a nearly massless scalar is necessary to drive the acceleration of the expansion of the Universe. We find here that its coupling to the Higgs field could result in a screening of its potentially induced fifth force in the solar system for instance. Applications of this mechanism to dark energy models are left for future work.

This mechanism applies to the relaxion model where the $\mu$ term is a linear function of a scalar field modulated by a cosine function. We find that vacua with massless excitations do exist for the relaxion models and that the flat direction can be stabilised at quadratic order. Hence the fifth force due to the massless field in the relaxion spectrum can be  screened by the non-linear screening mechanism that we describe in this paper, in particular the models whose flat direction is stable under quantum corrections are screened as the scalar mass at the  minimum of the effective potential along the flat direction is large.

The paper is arranged as follows. In  section \ref{sec:higgs} we describe the Higgs portal models and their flat directions. We also consider the quantum corrections and impose restrictions on the models from the quantum stability of the flat directions. In section \ref{sec:long}, we describe the long range fifth force in the context of the linear response theory. When this breaks down, i.e. when there is a flat direction associated to a zero mass eigen-state, we study the stabilisation of the flat direction in matter in section \ref{sec:break} together with the new screening mechanisms particular to the Higgs-portal systems. Finally in section \ref{sec:relaxion} we focus on the relaxion cases. Technical details can be found in appendices.

\section{A light scalar coupled through the Higgs portal}
\label{sec:higgs}

In this section we will introduce the two field model that we work with in this article, we will determine the expectation values and masses of the fields in different environments, and will determine when effective single field approximations to the dynamics of the theory are useful. We end with a discussion of quantum corrections.

\subsection{The model}

We consider a theory which contains a light scalar $\phi$ in addition to the Higgs field $h$ and a fermion field $\psi$.  We thus study a simplified version of the Standard Model, so that the Higgs field is real, and only one Dirac fermion is present.   We expect the generalisation of the results we derive here to the full Standard Model to be straight forward.  The Lagrangian we consider is
\be
{\cal L}= -\frac{1}{2} (\partial \phi)^2 -\frac{1}{2} (\partial h)^2 - V(\phi) + \frac{\mu^2(\phi)}{2} h^2 -\frac{\lambda}{4} h^4 - i \bar \psi \slashed{\partial} \psi - \lambda_\psi h \bar \psi \psi
\label{eq:higgsportal}
\ee
{ where the $\mu$ term is now field dependent.}
The scalar field $\phi$ could play the role of dark matter or dark energy. We therefore assume that the mass scales associated to the scalar $\phi$ are much lower than the mass of the Higgs field in vacuum. { As a result the lifetime of the scalar field is much larger than the age of the Universe \cite{Flacke:2016szy}.}
The corresponding equations of motion for the two scalars are
\begin{align}
\Box \phi -V^{\prime}(\phi) + \mu(\phi)\mu^{\prime}(\phi) h^2 &=0\\
\Box h +\mu^2(\phi) h -\lambda h^3 -\lambda_{\psi}\bar{\psi}\psi &=0
\end{align}
We denote the light scalar  vev by $\phi_0$ and the Higgs vev as $v$.  These must satisfy the following requirements
\begin{align}
\mu^2(\phi_0) &= \lambda v^2\\
V^{\prime}(\phi_0) &= - \frac{\mu^3 (\phi_0) \mu^{\prime}(\phi_0)}{\lambda}
\end{align}
In this work we will largely be interested in the behaviour of the fields around large bodies of dense matter, where we replace $\bar \psi \psi \to \langle \bar \psi \psi \rangle = n_\psi$ where $n_\psi$ is the number density of the medium. After identifying the mass of the fermions as $m_\psi=\lambda_\psi v$, the equations of motion for the fields in a dense medium become
\begin{align}
\Box \phi -V^{\prime}(\phi) + \mu(\phi)\mu^{\prime}(\phi) h^2 &=0\\
\Box h +\mu^2(\phi) h -\lambda h^3 -\frac{\rho}{v} &=0
\label{eqmot}
\end{align}
where $\rho$ is the local density of matter made up of our fermions $\psi$\footnote{Notice that we couple the Higgs field to fundamental fermions such as quarks and electrons. In a real material, most of the mass of the atoms comes from the masses of neutrons and protons. The mass of neutrons and protons comprises mostly the gluon part which is not coupled to the Higgs field and a small fraction due to the valence quarks. This introduces a proportionality coefficient $\alpha$ which can be extracted from \cite{Damour:1994zq,Brax:2006dc,Ahlers:2008qc,Bellazzini:2011et}. Hence the Higgs field should couple to $\alpha \rho$ in (\ref{eqmot}). We ignore this $\alpha$ coefficient in what follows. }
The resulting dynamics of the fields are controlled by an effective potential
\begin{equation}
V_{\rm eff}(h,\phi)= V(\phi) - \frac{1}{2}\mu^2(\phi) h^2 +\frac{\lambda}{4} h^4 +\frac{h \rho}{v}
\end{equation}
In what follows we will use the subscript $0$ to denote the expectation values of the fields in vacuum, and the subscript `bg' to denote the values of the fields which minimise the effective potential of the theory in a background of constant non-zero density $\rho$.

\subsection{An effective field theory for $\phi$}
\label{sec:efffield}

If there is a  large hierarchy between the mass of the Higgs and the mass of the light scalar field, then we might expect to be able to `integrate out' the Higgs field, leaving only an effective theory for $\phi$.
In vacuum, or when gradients of $h$ can be ignored, we can integrate out the Higgs field when heavy enough compared to the light scalar $\phi$  using the minimum equation along the $h$ direction, i.e. $\frac{\partial {\cal L}}{\partial h}=0$. This leads to
\be
\lambda h^2(\phi)= \mu^2 (\phi) - \frac{\lambda_\psi}{h(\phi)} \bar \psi \psi.
\ee
To first order in perturbations of $h$ around its vev, we  have
\be
 h(\phi)= \frac{\mu (\phi)}{\sqrt{\lambda}} - \frac{\lambda_\psi}{2  \mu^2 (\phi)} \bar \psi \psi
\ee
which gives the effective Lagrangian
\be
{\cal L}= -\frac{1}{2} (\partial \phi)^2 + \frac{\mu^4(\phi)}{4\lambda} -V(\phi) - i \bar \psi \slashed{\partial} \psi - \frac{\mu(\phi)}{\sqrt \lambda v} m_\psi  \bar \psi \psi
\ee
 and the  strength of the tree level scalar exchange between two fermions is controlled by the coupling constant
\begin{equation}
\frac{\beta_{\phi}(\phi)}{m_{\rm Pl}} = -\frac{\mu^{\prime}(\phi)}{\sqrt \lambda v}
\label{eq:effcoupling}
\end{equation}
corresponding to a Yukawa interaction mediated by perturbations in the field, $\delta\phi= \phi-\phi_0$, of the form
\be
{\cal L} \supset \frac{\beta_\phi}{m_{\rm Pl}} \delta\phi \rho
\ee

In a background of density $\rho= m_{\psi} \langle \bar \psi \psi \rangle$, the background value $\phi_{\rm bg}$ is determined by
\be
V_{\rm eff}^{\prime}(\phi_{\rm bg})= V^{\prime}(\phi_{\rm bg}) -\frac{\mu^{\prime}(\phi_{\rm bg}) \mu^3 (\phi_{\rm bg})}{\lambda}+ \frac{\mu^{\prime}(\phi_{\rm bg})\rho}{\sqrt{\lambda}v}=0
\ee
where the one-field effective potential $V_{\rm eff}(\phi)$ given by
\be
V_{\rm eff}(\phi)= V_{\rm eff}(h(\phi),\phi)
\ee
 determines the dynamics of the field $\phi$ in an environment of density $\rho$.

In the medium of density $\rho$, the scalar field acquires a mass given by
where
\be
m_{\phi,\rm bg}^2= -\frac{1}{\lambda}[ 3 \mu^2(\phi_{\rm bg})(\mu^{\prime}(\phi_{\rm bg}))^2 +\mu^3(\phi_{\rm bg})\mu^{\prime\prime}(\phi_{\rm bg})] + V^{\prime \prime}(\phi_{\rm bg})
\ee
and mediates a fifth force proportional to the Newtonian interaction
\be
V(r)= 2\beta_{\phi}^2(\phi_{\rm bg}) V_N(r) e^{-m_{\phi,\rm bg}r}
\label{eq:yukawapot}
\ee
where the Newtonian interaction between two fermions of masses $m_\psi$ is
\be
V_N(r)= -\frac{G_N m_\psi^2}{r}.
\ee
This is a  Yukawa interaction whose strength is determined by the coupling $\beta_\phi(\phi_{\rm bg})$. The force mediated by the scalar $\phi$ is weaker than gravity if $|\beta_\phi(\phi_{\rm bg})|<1/\sqrt 2 $.

If there is no scalar potential, i.e. $V(\phi)\equiv 0$, there exists an extremum of the  effective potential for the scalar field when
\be
\mu^{\prime}(\phi_0) = 0.
\ee
Without knowing  further details of the model we cannot say whether this is the true minimum of the theory.  If it is the  minimum, however, there is no scalar mediated fifth force because the coupling constant vanishes.  If the scalar has a bare potential $V(\phi)$ then the vacuum value of the scalar field is shifted, i.e. opening up the possibility of a  fifth force in vacuum which may still be screened in more dense environments. We will discuss when this is the case in particular if the Higgs-scalar mass matrix has a vanishing eigenmass, i.e. if there exists a potentially unscreened and long range scalar interaction.

\subsection{The effects of mixing}

In this section we return to the full two field model of equation (\ref{eq:higgsportal}) and will determine when the mixing between the two fields becomes so significant that the dynamics of the Higgs field cannot be neglected.
We start by considering the mass matrix for the theory in vacuum, this is made of the second derivatives of the potential in eq. (\ref{eq:higgsportal}).
We define
\begin{align}
m_h^2 &= 2 \mu^2(\phi_0)\\
m_{\phi}^2 &= V^{\prime\prime}(\phi_0) -v^2 (\mu^{\prime}(\phi_0))^2 -v^2 \mu(\phi_0)\mu^{\prime\prime}(\phi_0)
\end{align}
and  a vacuum mixing angle $\theta$ such that\footnote{Note that, in the vacuum, we have the link between the mixing angle and the coupling to matter defined in the effective theory for $\phi$ in equation (\ref{eq:effcoupling}),
\be
\frac{\beta_\phi}{m_{\rm Pl}}\equiv \frac{\beta_\phi (\phi_0)}{m_{\rm Pl}}= \frac{\sin \theta}{v}
\label{eq:singlebeta}
\ee
We will see this direct link reappear in Section \ref{sec:lin}.}
\begin{equation}
\sin \theta = \frac{1}{m_h^2}\left.\frac{\partial^2 V(\phi,h)}{\partial \phi \partial h}\right|_{\phi_0,v} = -\frac{v \mu^{\prime}(\phi_0)}{\mu(\phi_0)}.
\end{equation}

The mass matrix of the scalar potential is given by
\begin{equation}
\mathcal{M}^2 = \left(\begin{array}{cc}
V_{hh} & V_{h\phi}\\
V_{h\phi} & V_{\phi\phi}\\
\end{array}\right)
\label{eq:massmat}
\end{equation}
where $V_{xy}=\frac{\partial^2 V}{\partial x \partial y}$ are the second derivatives of the potential, and we identify  $x$ and $y$  either with $h$ or $\phi$. The mass matrix in vacuum becomes
\begin{equation}
\mathcal{M}_0^2 = \left(\begin{array}{cc}
m_h^2 & m_h^2 \sin\theta \\
m_h^2 \sin\theta & m_{\phi}^2
\end{array}\right)
\end{equation}
whose mass eigenvalues are
\begin{equation}
m_{\pm}^2 = \frac{1}{2}\left( m_h^2 +m_{\phi}^2 \pm \sqrt{(m_h^2-m_{\phi}^2)^2 +4 m_h^4\sin^2 \theta} \right)
\end{equation}
We will typically work in the regime where $m_{\phi} \ll m_h$ and $\theta \ll 1$ in which case, to first order in small quantities,  the two mass eigenvalues are $m_+^2\approx m_h^2$ and $m_-^2\approx m_{\phi}^2 -m_h^2\sin^2\theta$.
When the mixing angle is much smaller than $m_\phi/m_h$, the mass matrix is essentially diagonal with eigenmasses given by $m_h$ and $m_\phi$. As the mixing angle increases, the light mode of mass $m_-$ becomes lighter until
\be
\vert \sin \theta \vert = \frac{m_{\phi}}{m_h}
\ee
where the eigenmass $m_-=0$ vanishes.
We will see in Section \ref{sec:breakdown}, that when the mass $m_- \approx 0$ it is necessary to work to higher order in perturbation theory in the presence of matter.

To see why a linearised perturbative analysis fails,  consider how the fields behave in a dense environment. Assuming that a local matter density perturbs the fields from their vacuum expectation values, the background values of the fields satisfy
\be
h_{\rm bg}^2 = \frac{V^{\prime}(\phi_{\rm bg})}{\mu(\phi_{\rm bg}) \mu^{\prime}(\phi_{\rm bg})}
\label{eq:h1}
\ee
and
\be
h_{\rm bg} \left ( \mu^2(\phi_{\rm bg}) -\frac{\lambda V^{\prime}(\phi_{\rm bg})}{\mu(\phi_{\rm bg}) \mu^{\prime}(\phi_{\rm bg})}\right)  = \frac{\rho}{v} \label{eq:pertur}
\ee
If these field values are close to the values that minimise the effective potential in vacuum, we can write $\phi_{\rm bg}=\phi_0+\delta \phi$ and $h_{\rm bg}=v+\delta h$. The solution to these equation is given by
\begin{align}
\delta h &=  -\frac{m_{\phi}^2}{m_h^2}\frac{\delta \phi}{\sin \theta}\\
\delta \phi &= \frac{\beta_\phi \rho}{m_{\rm Pl}} \frac{1}{m_{\phi}^2-m_h^2 \sin^2 \theta }
\label{delmin}
\end{align}
where $\beta_\phi$ is the coupling in vacuum, i.e. $\beta_\phi=\frac{m_{\rm Pl}}{v}\sin\theta$.
Even in low density environments, we see that the linear perturbative treatment   fails when $m_\phi=|\sin\theta| m_h$, i.e. when $m_-=0$.

\subsection{Effective single field theory for the light mode}
\label{sec:breakdown}

The breakdown of linear perturbation theory   can be better understood by studying the theory along the nearly flat direction
defined by the  eigenvector associated to the eigenmass $m_-$ for $m_-/m_\phi \ll 1$.   i.e.
\be
\vec \alpha_-= \left ( \begin{array}{c}
-\sin\theta\\
1\\
\end{array}
\right )
\ee
 As in Section \ref{sec:efffield} we are reducing the theory to a single field effective model, but this time identifying that the true light mode of the theory is a mix of both the Higgs and the field $\phi$.
This amounts to analysing the potential along the field direction parameterised by the fields $(h,\phi)= ( v- \sin\theta \delta\phi, \phi_0 +\delta\phi)$.  The dynamics are governed by a potential
\be
U(\delta\phi)= V_{\rm eff}(v- \sin\theta \delta\phi, \phi_0 +\delta\phi)-V_{\rm eff}(v,\phi_0)
\ee
which  can be expanded to cubic order as
\be
U(\delta\phi)\simeq \frac{1}{2} m_-^2 \delta\phi^2 + \frac{1}{6}U^{\prime\prime\prime} \delta\phi^3 - \sin\theta \frac{\delta \phi}{v} \rho.
\label{eq:U}
\ee
where a prime indicates a derivative with respect to $\phi$. The third derivative, and coefficient of $\delta \phi^3$ in Eq. (\ref{eq:U}) is
\be
U^{\prime\prime\prime} = V_{\phi\phi\phi} -\frac{v^2}{2}\left( (\mu^2)^{\prime\prime\prime} +{3} \frac{(\mu^2)^\prime (\mu^2)^{\prime\prime}}{\mu^2}\right)
\ee
evaluated at $(h,\phi)=(v,\phi_0)$. The dashes are derivatives with respect to $\phi$.

Neglecting the cubic term, the minimum of the potential in equation (\ref{eq:U}) is obtained for
\begin{align}
\delta \phi &=  \frac{\sin\theta}{m_-^2} \frac{\rho}{v}
\end{align}
which coincides  with Eq. (\ref{delmin}), the position of the minimum of the potential for the linearised two field theory,  when $\sin\theta \simeq \frac{m_{\phi}}{m_h}$. We  now obtain a  criterion for the breakdown of linear perturbation theory; higher order perturbations must be included  when the cubic terms dominates over the mass term, i.e. for mass eigenvalues such that
\be
m_-^2 \lesssim \left \vert \frac{U^{\prime\prime\prime}}{v}  \right \vert \sin \theta  {\rho}
\label{break}
\ee
We will illustrate this breakdown of linear perturbation theory for the relaxion model, when it is introduced in Section \ref{sec:relaxion}.

When the cubic term in the effective potential becomes important, the equation for the position of the minimum of the potential is modified. A new minimum can be found when
\be
U^{\prime\prime\prime} >0.
\ee
and the field is stabilised with a vev
\be
\delta \phi_{\rm bg} =  \sqrt{\frac{2m_\phi \rho }{U^{\prime\prime\prime}  m_h v}}
\ee
where the square of the mass of fluctuations of the field at the position of the  new minimum is
\be
m^2_{-\rm bg}= U^{\prime\prime\prime}  \delta  \phi_{\rm bg}.
\label{eq:Umass}
\ee
We show in the appendix \ref{app:B} that the mass  given in Eq. (\ref{eq:Umass})
and the lowest eigenvalue
of the Higgs-scalar mass matrix coincide.
When $U^{\prime\prime\prime}<0 $, the flat direction is not stabilised at the quadratic order. Stabilization at higher order would require  unnatural fine-tunings of the coefficients of the perturbation series of the full
potential around the vacuum expectation values, or for the theory to evolve into a fully non-perturbative regime.
We will analyse the fate of these models briefly in the appendix \ref{app:D}.

\subsection{Quantum corrections}

As we are interested in a nearly flat direction in field space, we should worry about the quantum stability of that flat part of the potential. A theory extending  the Standard Model with a second scalar field coupled through the Higgs portal   suffers from the usual hierarchy problem of the standard model. This is the question of the stability of the Higgs vev and mass under quantum corrections induced by heavy, i.e. larger than the Higgs mass,  states\footnote{The relaxion model \cite{Graham:2015cka} is an exception to this, where the second scalar field is introduced to try to provide a solution for the hierarchy problem.  We will discuss the relaxion further in Section \ref{sec:relaxion}.} In the context of our model this becomes a question about the sensitivity of the potential $V(\phi)$ and the mass function $\mu(\phi)$ to corrections from quantum fluctuations of beyond the standard model particles. We have nothing to add to this discussion and take for granted that the low energy effective action for the Higgs-scalar system does admit a flat direction when the mixing angle is close to $m_\phi/m_h$. However, we must  check that at low energies, below the electro-weak scale, the dynamics of the field $\delta\phi$, parameterising the flat direction and stabilised by dense matter, is not dominated  by quantum corrections.

We work with  the effective single field potential described in Section \ref{sec:breakdown}, where  $\delta\phi$ describes fluctuations of the light mode around the minimum of the effective potential. The resulting quantum corrections  depend on the mass of these fluctuations given by
\be
m^2_{\delta\phi}= m_-^2 + U^{\prime\prime\prime} \delta\phi.
\label{eq:lightmass}
\ee
The one loop correction to the scalar potential calculated in dimensional regularisation with a sliding scale $\mu_{\rm DR}$ is\footnote{We regularise divergent integrals like
$\int \frac{ d^4 p}{p^2+m^2(\delta\phi)}= m^2(\delta\phi) (\frac{\mu}{m_-})^{\epsilon_{\rm DR}} \int\frac{d^{4+\epsilon_{\rm DR}} x}{x^2+1}$ and extract the finite part, depending on $\ln \mu/m_-$, after removing the divergences in $1/\epsilon_{\rm DR}$.}
\be
\delta_\phi U(\delta\phi)= -\frac{m^4(\delta \phi)}{64\pi^2} \ln \frac{\mu^2_{\rm DR}}{m^2_-}.
\ee
When $m_-$ is negligible in Eq. (\ref{eq:lightmass}), the correction to the potential becomes
\be
\delta_\phi U(\delta\phi)\propto  -\frac{(U^{\prime\prime\prime})^2  \delta\phi^2}{64\pi^2}.
\ee
to the mass term which does not lift the flat direction as long as
\be
 U^{\prime\prime\prime}\lesssim \frac{32 \pi^2}{3} v \epsilon.
 \label{quantum}
\ee
This constrains the parameter space of the Higgs-portal models with a flat direction.
In a similar way we can determine the effects of  quantum fluctuations of a massive fermion $\psi$, whose mass $m_\psi(\delta\phi)= m_\psi(1- \sin\theta \frac{\delta\phi}{v})$ comes from the variation of the Higgs field along the flat direction. Fluctuations of $\psi$ give a correction to the potential of the form
\be
\delta_\psi U(\delta\phi)= \frac{m^4_\psi (\delta \phi)}{32\pi^2} \ln \frac{\mu^2_{\rm DR}}{m^2_\psi}.
\ee

The full form of the potential along the flat direction, taking quantum corrections into account, is
\be
U(\delta \phi)= U_{\rm vac} + (m^2_\phi -\sin^2 \theta m^2_h) \frac{\delta\phi^2}{2} +  U^{\prime\prime\prime}\frac{\delta\phi^3}{6} + T \frac{\sin\theta \delta\phi}{v}
\label{eq:fullU}
\ee
where
$T$ is the trace of the energy momentum tensor  of matter and vacuum energy
\be
T= -\rho - 4U_{\rm vac}.
\ee
Notice that the quantum corrections due to the fermions $\psi$ to the vacuum energy induce the change
\be
U_{\rm vac} \to U_{\rm vac} + \frac{m^4_\psi }{32\pi^2} \ln \frac{\mu^2_{\rm DR}}{m^2_\psi}
\ee
and this can be renormalised to zero in what follows.

The quantum corrections due to the fermion fluctuations do not modify the  term cubic in $\delta\phi$ in Eq. (\ref{eq:fullU}) as long as
\be
U^{\prime\prime\prime}\gtrsim \sin^3\theta m_\psi
\ee
The quadratic correction to the mass term amounts to changing
\be
m^2_h \to m_h^2 - \frac{3m^4_\psi}{8\pi^2 v^2} \ln \frac{\mu^2_{\rm DR}}{m^2_\psi}
\ee
which is the quantum correction to the Higgs mass due to the fermion loop\footnote{The correction to the mass of the Higgs due to the top quark is given in Higgs Hunter's Guide~\cite{Gunion:1989we} page 66 and reads $m^2_h \to m^2_h -\frac{3m_t^4}{2\pi^2 v^2}\ln \frac{\mu^2}{Q^2}$ where $\mu^2= v^2/2$ is their renormalisation scale and the dependence on the energy $Q$ comes from the rescaling $\mu\to \mu/Q$ in scattering amplitudes. As the complex Higgs field takes a vev $v/\sqrt 2 $, the top mass is related to the $\psi$ mass as $m_\psi= \sqrt 2 m_t$. This completes the identification.}. Again, this can be absorbed in the definition of the Higgs mass.
In the end, the quantum corrected potential is similar to the classical one after renormalisation of the Higgs mass and the vacuum energy. Higher order terms in $\delta\phi^4$ are negligible due to the $\sin^4\theta$ factor. Hence we find that the flat direction is preserved quantum mechanically. In the following, we will assume that the Higgs mass is the renormalised one and analyse the flat direction for $\sin\theta= \frac{m_\phi}{m_h}$ where $m_h$ has been renormalised.

\section{The long range fifth force}
\label{sec:long}
So far  we have discussed the values of the two scalar field  that minimise the effective potential in vacuum and in dense environments, and the resulting effective coupling of the light mode to matter.
In this section we will proceed to analyse the form of the scalar fifth force around a compact object more precisely.\footnote{We use the term `fifth force' here to denote the new long range scalar mediated force, that is not present in the Standard Model.  However, the Higgs, as a scalar field, in principle mediates a very short range force that could be termed the `fifth force'. In which case the additional scalar field is actually mediating a `sixth force'.}
Around a compact matter source the two fields will be perturbed from their background values. As the mass eigenstates are combinations of the Higgs field and scalar $\phi$ the perturbation to the Higgs field will contain a component of the light mass eigenstate, the gradient of which could communicate a long range fifth force to other matter particles in the vicinity. This is the way a compact object interacts with the surrounding matter particles with a range characterised by the Compton wavelength associated with the lowest eigenmass of the Higgs-scalar mass matrix in vacuum.

\subsection{Spherical profiles}
 To study the behaviour of the fifth force we will assume a source mass made of matter which is  a static sphere of constant density $\rho$ and radius $R$ (so that the total mass of the source is $M=4 \pi \rho R^3 /3$).  We will solve the equations of motion for the fields both inside and outside the source mass and impose continuity of the fields and their derivatives at the surface.

Both inside and outside the source the mass matrix, Eq. (\ref{eq:massmat}), can be written in terms of a diagonal matrix of the eigenvalues, ${\cal M}_D^2$ and a rotation matrix, $R$, as  $\mathcal{M}^2 = R^{T}{\cal M}^2_DR$, where the superscript $T$ denotes the transpose of the matrix, see appendix \ref{app:B}.   We allow for the mass matrix to be different inside and outside the source, we will say that outside the source the  eigenvalues of $\mathcal{M}^2$ are $y_{\pm}$, and inside the source these are $\lambda_{\pm}$ (both $y_{\pm}$ and $\lambda_{\pm}$ have dimensions of mass squared). The rotation matrices are $R_i$ and $R_o$, which can be expressed in terms of mixing angles $\theta_i$ and $\theta_o$, see appendix (\ref{app:B}).

The equations of motion to be solved are therefore
\begin{equation}
  R_o \Delta\left(\begin{array}{c}
h- v\\
 \phi -\phi_0
\end{array}\right)
= \left( \begin{array}{cc}
y_+ & 0 \\
0 & y_-
\end{array}\right) R_o \left(
\begin{array}{c}
 h - v\\
 \phi- \phi_0
\end{array}\right)
\label{eq:eomout}
\end{equation}
outside the source, and
\begin{equation}
 R_i \Delta\left(\begin{array}{c}
 h-h_{\rm in}\\
\phi - \phi_{\rm in}
\end{array}\right)
= \left( \begin{array}{cc}
\lambda_+ & 0 \\
0 & \lambda_-
\end{array}\right) R_i \left(
\begin{array}{c}
h-h_{\rm in}\\
\phi - \phi_{\rm in}
\end{array}\right)
\label{eq:eomin}
\end{equation}
inside the source, where  $h_{\rm in}$ and $\phi_{\rm in}$  are the values of the fields that minimise the effective potential inside the source and $\Delta$ the Laplacian operator in spherical coordinates.

\subsection{The linear case}
\label{sec:lin}

We begin by studying the case where linear perturbation theory can be trusted.  This assumes that the model we study is far away, in parameter space, from points where $m_{\phi}^2 = m_h^2 \sin^2 \theta$, and that the density of the source is not too high; see for instance Eq. (\ref{break}). In this case we know that the values of the fields that minimise the effective potential inside the source are small perturbations from the values that minimise the effective potential in vacuum. In the linear case, the mass matrices can be considered to be equal, i.e. the mixing angles $\theta_{i,o}$ coincide inside and outside the compact objects. This will not be the case in the quadratic case as we will see next.

As a result, the equation of motion inside the source, Eq. (\ref{eq:eomin}), can be simplified to be
\begin{equation}
 R_o \Delta \left(\begin{array}{c}
h-v\\
\phi - \phi_0
\end{array}\right)
= \left( \begin{array}{cc}
y_+ & 0 \\
0 & y_-
\end{array}\right) R_o \left(
\begin{array}{c}
h-v\\
\phi - \phi_0
\end{array}\right)
+\frac{\rho}{v} R_o \left(
\begin{array}{c}
1\\
0
\end{array}\right)
\label{eq:eomlinin}
\end{equation}
Notice that we have explicitly use the equality between the mass matrices inside and outside the source.

The equations of motion in  Eq. (\ref{eq:eomlinin}) can be solved in terms of the rotated fields
\begin{equation}
\delta \Phi = R_o \left(\begin{array}{c}
h-v\\
\phi - \phi_0
\end{array}\right)
\end{equation}
which must be identified with decreasing exponential functions outside the source
\begin{equation}
\delta \Phi_o = \left(\begin{array}{c}
\frac{B_+}{\sqrt{y_+}r}e^{-\sqrt{y_+}(r-R_0)}\\
\frac{B_-}{\sqrt{y_-}r}e^{-\sqrt{y_-}(r-R_0)}
\end{array}\right).
\end{equation}
where $B_{\pm}$ are constants of integration.
Inside the source one must impose that the first derivatives of the fields must vanish at the origin leading to
\begin{equation}
\delta \Phi_i = \left(\begin{array}{c}
\frac{A_+}{\sqrt{y_+}r}\sinh\sqrt{y_+}r\\
\frac{A_-}{\sqrt{y_-} r}\sinh\sqrt{y_-}r
\end{array}\right)
-\left(\begin{array}{c}
\frac{C}{y_+}\\
\frac{D}{y_-}
\end{array}\right)
\end{equation}
where
\begin{equation}
\left(\begin{array}{c}
C\\
D
\end{array}\right) = \frac{\rho}{v} R_o \left(\begin{array}{c}
1\\
0
\end{array}\right)
\end{equation}
The constants $A_{\pm}$ and $B_{\pm}$ are constants of integration. As already stated, we have imposed the boundary conditions that the perturbations to the fields should decay to zero as $r \rightarrow \infty$ and that the fields should be regular at $r=0$.

By requiring that the fields and their first derivatives are continuous at the surface of the source we determine the remaining constants of integration.  The full expressions for these constants are lengthy algebraic expressions, see  section  \ref{sec:nonlin} and appendix \ref{app:C} for the complete expressions.
For simplicity  we focus on the physically interesting case where the Compton wavelength of the heavy mode is always smaller than the size of the source, and the Compton wavelength of the light mode is always larger.  This requires the assumptions:
\begin{align}
\sqrt{y_+}R \gg 1\;, & \sqrt{y_-}R\ll 1\;
\end{align}
and corresponds to the physical situation of a Higgs field mediating an interaction of sub-Fermi range and a light scalar with a Compton wavelength much greater than the size of the source.
Of greatest relevance  is the way in which the light mode appears in the Higgs field.  Outside the source this appears in  the form
\begin{equation}
h \approx v -\frac{\sin\theta_o B_-}{\sqrt{y_-} r} e^{-\sqrt{y_-}(r-R)}
\end{equation}
This will communicate a long range interaction to the matter fields depending on the mixing constant $B_-$.

Keeping only the leading terms, we find that the four constants of integration are given by
\begin{align}
\frac{B_+}{\sqrt{y_+}} & = -\frac{\rho \cos \theta_o R}{2 y_+ v}\\
\frac{B_-}{\sqrt{y_-}} & = \frac{\rho \sin\theta_o R^3}{3 v}\\
\frac{A_+}{\sqrt{y_+}} e^{\sqrt{y_+}R} & = \frac{\rho \cos\theta_o R}{v y_+}\\
\frac{A_-}{\sqrt{y_-}} & =-\frac{-\rho \sin\theta_o}{v y_-^{3/2}}
\end{align}
The tree level long range scalar force is communicated to matter  through the Higgs field, which has the form,
\begin{equation}
h \approx v -\frac{M\sin^2\theta_o }{4 \pi v r} e^{-\sqrt{y_-}(r-R)}
\label{eq:lin}
\end{equation}
where $M= (4 \pi / 3) \rho R^3$, as expected in the linear theory (recall that in this case $y_-\approx m_{\phi}^2 -m_h^2 \sin^2 \theta_o$).
Thus the light mode, of mass $\sqrt{y_-}$, mediates a Yukawa interaction between matter particles and massive objects, with a coupling strength relative to gravity
\be
\frac{\beta_\phi}{m_{\rm Pl}} = \frac{ \sin\theta_o }{v}
\ee
This is in exact agreement with the coupling constant we determined for the  effective field theory where the Higgs had been integrated out in Eq. (\ref{eq:singlebeta}).  So we can conclude that the effective single field theory for $\phi$ accurately  captures the behaviour of the fifth force in the linear regime.

\section{The Higgs portal screening mechanism}
\label{sec:break}
\subsection{The perturbed fields}
\label{sec:nonlin}
In the previous section, we showed that, when  linear perturbation theory  is valid,   the strength of the interaction is proportional to $\beta^2_\phi$.
We will now show how this  behaviour is modified in the non-linear case when the Higgs-scalar mass matrix admits a flat direction. In this case, linear perturbation theory breaks down as
\begin{equation}
m_h^2 \sin^2 \theta = m_{\phi}^2.
\end{equation}
when the mass of the lightest mode (almost) vanishes.  This does not mean that the system is no longer perturbative.
Indeed the  flat direction in field space  is lifted  as long as $U^{\prime\prime\prime}>0$ at second order in perturbation theory.
These minimum values for the two fields are given by
\be
\phi_{\rm in} = \phi_0 + v \epsilon
\label{nonlinear}
\ee
\be
h_{\rm in}=v\left(1- \frac{m_\phi}{m_h } \epsilon \right)
\label{eq:nonlinh}
\ee
where we have introduced the dimensionless parameter
\be
\epsilon= \sqrt{\frac{2m_\phi \rho }{U^{\prime\prime\prime}  m_h v^3}}
\ee
The perturbations to $\phi_0$ and $v$ will remain small, due to the smallness of $\rho$ compared to the particle physics scales, unless $U^{\prime\prime\prime}$ is vanishingly tiny; we will assume in what follows that this is not the case and will analyse the range of values of $U^{\prime\prime\prime}$ allowed by observations below.
The displacements in both $\phi$ and $h$ due to the presence of matter with density $\rho$ are now proportional to $\sqrt{\rho}$.  The fields react less strongly to the presence of a dense source, i.e. in $\sqrt \rho$,  than they do in the linear regime, i.e. in $\rho$.  So we start to see screening of the effects of the scalar fields emerging.

 The mass matrix in a region of density $\rho$  has eigenvalues
\begin{align}
\lambda_+ & \approx m_h^2\\
\lambda_- & \approx  U^{\prime\prime\prime} \epsilon v
\end{align}
and the mixing angle is
\begin{equation}
\theta_i \approx \theta_o -\delta
\end{equation}
where we have a small mixing angle
\be
\theta_o\simeq \frac{m_{\phi}}{m_h}
\ee
and the variation of this mixing angle due to the presence of matter is given
\be
\delta= \left( \frac{m^2_\phi}{m^2_h}- \frac{(\mu^2)^{\prime\prime}(\phi_0)}{2\lambda}\right) \epsilon.
\ee
To find the form of the long range fifth force we solve the equations of motion for the fields in the same way as we outlined in the previous section for the linear case. We find that
outside the source
\begin{equation}
\left(\begin{array}{c}
h\\
\phi
\end{array}\right)
= \left(\begin{array}{c}
v\\
\phi_0
\end{array}\right)
+R^T_o
\left(\begin{array}{c}
\frac{B_+}{\sqrt{y_+}r}e^{-\sqrt{y_+}(r-R_0)}\\
\frac{B_-}{\sqrt{y_-}r}e^{-\sqrt{y_-}(r-R_0)}
\end{array}\right)
\end{equation}
and inside the source
\begin{equation}
\left(\begin{array}{c}
h\\
\phi
\end{array}\right)
= \left(\begin{array}{c}
h_{\rm in}\\
\phi_{\rm in}
\end{array}\right)
+R^T_i
 \left(\begin{array}{c}
\frac{A_+}{\sqrt{\lambda_+}r}\sinh\sqrt{\lambda_+}r\\
\frac{A_-}{\sqrt{\lambda_-}r}\sinh\sqrt{\lambda_-}r
\end{array}\right)
\end{equation}
again, $A_{\pm}$ and $B_{\pm}$ are constants of integration and we have imposed the boundary conditions that the perturbations to the fields should decay to zero as $r \rightarrow \infty$ and that the fields should be regular at $r=0$.

Imposing that the fields and their first derivatives are continuous at the surface of the source we find that the constants of integration are given by the following identities
\be
\left ( \begin{array} {c}
A_+\\
A_-
\end{array} \right)=- X^{-1} (1+Y) R_o \delta\Phi
\ee
and
\be
\left ( \begin{array} {c}
\frac{B_+}{\sqrt{y_+}R}\\
\frac{B_-}{\sqrt{y_-}R}
\end{array} \right)=(1-R_\delta S X^{-1} (1+Y)) R_o \delta \Phi
\label{ssol}
\ee
where we have introduced the vector
\be
\delta\Phi= \left ( \begin{array} {c}
\delta h\\
\delta \phi
\end{array} \right)
=\left ( \begin{array} {c}
h_{\rm in}-v\\
\phi_{\rm in} - \phi_0
\end{array} \right)
\ee
and the matrix $R_\delta =R_o R_i^T$, corresponding to the rotation angle
$
\delta = \theta_{o}-\theta_i .
$
We also define the matrices
\begin{eqnarray}
&& Y =\left ( \begin{array} {cc}
\sqrt y_+ R& 0 \\
0 &\sqrt y_- R
\end{array} \right),  \ \
C =\left ( \begin{array} {cc}
\cosh \sqrt{\lambda_+} R& 0 \\
0 &\cosh \sqrt{\lambda_-} R
\end{array} \right),\nonumber \\
&&S =\left ( \begin{array} {cc}
\frac{\sinh \sqrt{\lambda_+} R}{\sqrt \lambda_+ R}& 0 \\
0 &\frac{\sinh \sqrt{\lambda_-} R}{\sqrt \lambda_- R}
\end{array} \right), \ \
X= R_\delta C + Y R_\delta S. \nonumber \\
\end{eqnarray}
Notice that in the limit $y_-\to 0$, i.e. in the massless case, the constant $\frac{B_-}{\sqrt{y_-}R}$ has a finite limit.
Again we consider the case of most physical interest, where the Compton wavelength of the heavy mode is much smaller than the radius of the source and the Compton wavelength of the light mode is much larger;  $\sqrt\lambda_+ R\gg 1$ and $\sqrt\lambda_- R\ll 1$.  We also assume  $\lambda_+\simeq y_+\simeq m_h,\ \lambda_-\simeq y_-$. In this limit the relevant parameter $B_-$ simplifies and is given in the appendix \ref{app:C} where the full expression with no approximation can also be found.

\subsection{Screening nearly massless fields}
\subsubsection{The screening factor}

We now concentrate on the constant $B_-/\sqrt{y_-}$ when $y_- $ is vanishing small, i.e. when there is a nearly massless field in the spectrum of the theory. We also rely on the fact   that
the quadratic term of the effective potential $U(\delta\phi)$ is negligible compared to the cubic term, which stabilises the flat direction at quadratic order in the presence of matter. The scalar force outside the massive object is mediated by the Higgs field which couples to matter. In particular,
the screening of the scalar interaction compared to the linear case  depends on the profile of the Higgs field outside the objects. This depends on the $B_-$ coefficient which is evaluated in appendix (\ref{app:C}) and reads
\be
\frac{B_-}{{\sqrt y_-} }= \kappa R^3 \delta \phi
\ee
where $\lambda_-= m_-^2 $ is the square of the lowest eigenmass of the Higgs-scalar mass matrix in the dense body, and we have introduced the prefactor
\be
\kappa= \frac{\frac{ m_-^2}{3}+  \frac{\delta^2 m_h  }{2R}}{1 +  \frac{m_h R \delta^2}{2}}.
\ee
The resulting Higgs field outside the body is given by
\be
h= v-  \kappa \delta \phi\frac{R^3 \sin\theta_o}{r}
\ee
in the nearly massless limit where the exponential Yukawa suppression is negligible.
The corresponding force on a test body of mass $m_{\rm test}$ is simply
\be
F_\phi= -\kappa\frac{\delta \phi}{v}\frac{m_{\rm test} R^3\sin\theta_o}{r^2}
\ee
This is the main result of this paper.

Notice first that the scalar force is always attractive as $\kappa >0$. Moreover, one can easily retrieve the linear case
by neglecting  terms in $\delta^2$ and upon using $\delta \phi=  \frac{\sin \theta_o\rho}{m_-^2 v}$ we get
\be
F_\phi^{\rm linear}= - \frac{2\sin\theta_o^2 m^2_{\rm Pl}}{v^2} \frac{G_N M m_{\rm test}}{r^2}
\ee
corresponding to a ratio with the Newtonian interaction of $2 \beta_\phi^2$ where $\beta_\phi= \frac{m_{\rm Pl}}{v} \sin\theta_o$.

In the non-linear case, the scalar force is related to the linear force by
\be
F_\phi^{\rm nl}= \frac{3 \kappa   v \delta \phi}{\rho \sin\theta_o} F_\phi^{\rm linear}.
\ee
The coefficient
\be
\Theta= \frac{3 \kappa   v \delta \phi}{\rho \sin\theta_o}
\ee
is the screening factor. Screened scalar interactions correspond to $\Theta<1$.
We focus on two interesting cases.
\subsubsection{The large radius case  $R\gtrsim \delta^2  \frac{m_h}{m_-^2}$}

The screening factor is also given by
\be
\Theta= \frac{3 \kappa   v^2 \epsilon }{\rho \sin\theta_o}
\ee
and  we find that
\be
\Theta=\frac{2}{1 +  \frac{m_h R \delta^2}{2}}
\ee
implying for small objects, the scalar interaction is enhanced by a factor of two compared to the linear case. This is {\it anti-screening} { whose origin can be traced back directly to the non-linear behaviour of the scalar field along the flat direction.
Indeed assume that the leading correction along the flat direction is not cubic but of the type
\be
U(\delta\phi)\supset \frac{U^{(n)}}{n!}\delta\phi^n.
\label{nth}
\ee
The minimum induced by matter is now located at
\be
\delta\phi_{\rm bg}= \epsilon v
\ee
where
\be
{\epsilon= \left(\frac{(n-1)! \sin\theta \rho}{U^{(n)} v^n}\right)^{1/(n-1)}.}
\ee
The mass is  given by
\be
m_-^2= \frac{U^{(n)}}{(n-2)!}\delta\phi_{\rm bg}^{n-2}
\ee
leading to the screening factor
\be
\Theta=\frac{n-1}{1 +  \frac{m_h R \delta^2}{2}}.
\ee
As can be seen, this anti-screening originates from the number of way of connecting the field $\delta\phi$ from a given body
to another $\delta\phi$ associated to another body via the n-th order interaction {of Eq.~}(\ref{nth}). Indeed there are $(n-1)$ way of choosing the second
leg $\delta\phi$ once the first one has been chosen in a vertex of order $n$. }

When $n=3$, screening only occurs  when $R$ is larger than the screening radius defined by
\be
R_{\rm scr}= 4 (\delta^2 m_h)^{-1}
\ee
the screening factor becomes small and equal to
\be
\Theta \simeq \frac{ R_{\rm scr}}{ R} \ll 1.
\ee
In this case the scalar interaction is suppressed for objects much bigger than the screening radius.

\subsubsection{The small radius case $R\lesssim \delta^2  \frac{m_h}{m_-^2}$}
In the small radius  case, we find that
\be
\kappa= \frac{\frac{\delta^2 m_h  }{2}}{1 +  \frac{m_h R \delta^2}{2}}.
\ee
As the numerator is larger than the term leading to anti-screening by a factor of two, screening only occurs for radius larger than the screening radius. In this case we find that
\be
\Theta= \frac{v^2}{m^2_{\rm Pl}} \frac{\epsilon }{ 2\Phi_N \sin \theta_o}.
\ee
Hence screening happens for objects with a large enough Newtonian potential $\Phi_N= G_N M/R$ satisfying
\be
\Phi_N \gtrsim \frac{v^2}{m^2_{\rm Pl}} \frac{\epsilon }{ 2\sin \theta_o}.
\ee
As $v^2/m_{\rm Pl}^2\ll 1$ and $\sin\theta_0 \ll {1}$, the order of magnitude of $\Phi_N$ for screening to happen is left undetermined. But there will certainly be regions of parameter space for which the fifth force is suppressed.
This second type of screening is similar to the chameleon screening mechanism which takes place for objects with a large enough surface Newton potential. Here screening happens only for large enough objects with a large enough Newton potential at their surface.

\subsubsection{The effective coupling and the violation of the equivalence principle}

The most stringent tests of the modified Newton law has been performed by the Cassini probe \cite{Bertotti:2003rm}. In the environment of the Sun, the correction to the motion of a test satellite must be smaller than $10^{-5}$, i.e.\footnote{ Notice that in this section we do {not} take into account the fact that the Higgs field, and as a result the scalar $\phi$, do not couple to neutrons and protons but to the constituent quarks. This reduces the direct coupling $\beta_\phi$ by a factor $\alpha$ \textcolor{red}{\cite{Damour:1994zq,Brax:2006dc,Ahlers:2008qc,Bellazzini:2011et}} which depends on the type of material considered in each experimental set-up, e.g. the type of metals used to test the equivalence principle. This would relax the bounds quoted in this section accordingly.}
\be
2\Theta_\odot \beta_\phi^2 \lesssim \epsilon_{\rm cas}\simeq 2\cdot 10^{-5}.
\ee
Modifications of Newton's law in the laboratory have been probed down to a distance $d\simeq 0.1 \mbox{ mm}$. This also imposes strong constraints on models.

Objects of density $\rho$ and size $R$ couple to the long range scalar field with an effective coupling
\be
\beta_{\rm eff}= \Theta \beta_{\phi}
\ee
corresponding to a rescaling of Newton's constant as
\be
G_N\to (1+2 \beta_\phi \beta_{\rm eff})G_N
\ee
in the interaction described previously between a test particle coupled with the strength $\beta_\phi$ and the body with a strength $\beta_{\rm eff}$. Unscreened bodies have $\Theta=1$.
More generally two bodies characterised by screened couplings $\beta_{A,B}= \Theta_{A,B} \beta_\phi$ interact with a modified strength
\be
G_N \to (1+2\beta_A \beta_B) G_N.
\ee
As a result of the density and size dependence of the factors $\Theta_{A,B}$, two screened bodies fall differently in the gravitational and scalar fields created by another body $C$. The E\"otvos parameter is therefore
\be
\eta_{AB}= \frac{\vert \vec a_A -\vec a_B\vert}{\vert \vec a_A + \vec a_B \vert} \simeq \beta_\phi^2\Theta_C \vert \Theta_A- \Theta_B\vert.
\ee
There are two potentially stringent tests of the equivalence principle. The first one is the Lunar Laser Ranging experiment \cite{Williams:2012nc} where the acceleration of the Moon and the Earth in the gravitational field of the Sun are monitored. The second one is the Microscope satellite experiment \cite{Berge:2017ovy}  where two cylinders of different compositions fall in the gravitational field of the Earth. In both cases the E\"otvos parameter is constrained at the $10^{-13}$ and $10^{-14}$ level respectively.

\section{Screening the relaxion}

\label{sec:relaxion}
\subsection{The relaxion}

The relaxion model \cite{Graham:2015cka} is a theory where an additional scalar is added to the Standard Model.  This scalar couples to the standard model fields through the Higgs portal in such a way that the dynamics of the relaxion could provide an explanation for the hierarchy problem.
We work with the relaxion model of Ref. \cite{Flacke:2016szy} where
\begin{align}
V (\phi)& = -r g \Lambda^3 \phi\\
\mu^2(\phi) & = 2( \Lambda^2-g\Lambda \phi) +4 \frac{\Lambda_{\rm br}^4}{v^2} \cos \frac{\phi}{f}
\end{align}
and $\Lambda$ is the UV cut-off \footnote{We have changed the sign of the linear term with no physical consequences. In particular, with this choice the mixing angle $\sin\theta$ is positive when $r>0$ complying with the convention used in the rest of the paper.}.

\begin{figure*}
\begin{center}
\epsfxsize=9 cm \epsfysize=9 cm {\epsfbox{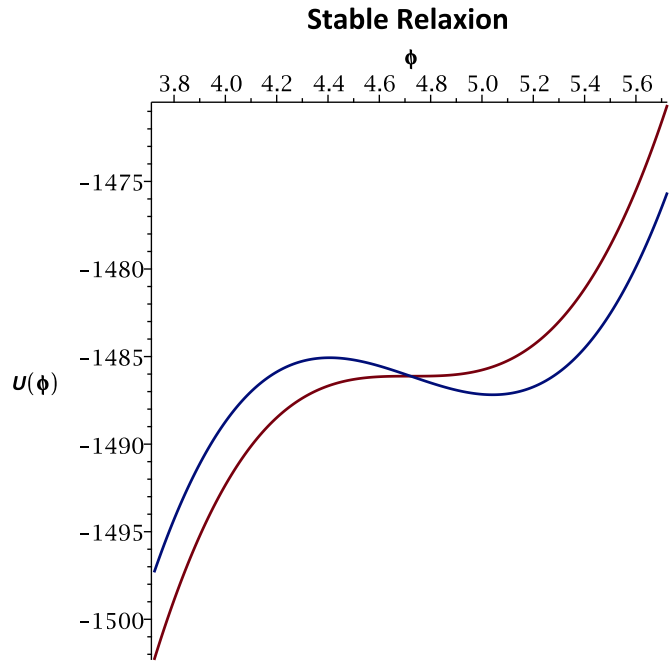}}
\end{center}
\caption{The stable potential along the flat direction for a relaxion model defined by  $\epsilon_0=0.01$ and $N=100$. The potential with a flat direction (mauve) is perturbed by matter (blue) and develops a minimum  in field space which depends on the square root of the matter density. }
\label{fig1}
\end{figure*}

The effective potential has a minimum in vacuum when the fields take the values $\phi_0$ and $v$ satisfying,
\begin{align}
v(-\Lambda^2 +g\Lambda\phi_0) + \frac{\lambda}{2} v^3 -\frac{2\Lambda^4_{\rm br}}{v}\cos\frac{\phi_0}{f} &=0\label{eq:v}\\
g\Lambda v^2 - rg\Lambda^3 +\frac{2\Lambda_{\rm br}^4}{f}\sin \frac{\phi_0}{f} &= 0\label{eq:phi0}
\end{align}
and the scalar masses are
\begin{align}
m_h^2 &= 2\lambda v^2\\
m_{\phi}^2 &= \frac{2 \Lambda_{\rm br}^4}{f^2} \cos \frac{\phi_0}{f}
\end{align}
The mixing angle is determined by
\be
\sin \theta= \frac{rg\Lambda^3}{\lambda v^3}.
\ee
The cubic parameter of the effective potential is given by
\be
U^{\prime \prime \prime}= -\frac{m^2_\phi}{f}\left(\tan \frac{\phi_0}{f} +6 \frac{f}{v} \sin\theta\right)
\ee
The light mode of the linearised theory is massless when
\begin{equation}
\cos \frac{\phi_0}{f} = \frac{g^2 r^2}{\lambda}\frac{f^2 \Lambda^6}{v^4 \Lambda_{\rm br}^4}
\end{equation}
and the flat direction is lifted at the quadratic order in the presence of matter when
\be
U^{\prime \prime \prime}= -\frac{m^2_\phi}{f}\left(\tan \frac{\phi_0}{f} +\frac{6}{N}\right)
\ee
is positive.
We have introduced the parameter
\be
N= \frac{vm_h}{fm_\phi}.
\label{defM}
\ee
The quadratic term of the effective potential is dominated by the cubic interaction when the density satisfies
\be
\rho > \frac{N^2}{2\vert 6+ N \tan\phi_0\vert} \frac{f^2 (m^2_\phi- \sin^2\theta m_h^2)^2}{m^2_\phi}.
\ee
In this region of parameter space of the relaxion, the screening mechanism can be at play.   We note that the role of higher order terms in the expansion of the relaxion potential have also been explored in Ref.~\cite{Banerjee:2020kww}.

\subsection{A generalised relaxion}

The analysis of the dynamics of the relaxion models along the flat direction can be simplified by writing the effective potential as
\begin{align}
V_{\rm eff}(h,\phi) = m_{\phi}^2 f^2&\left[\frac{N^2h^2}{8v^2}\left(\frac{h^2}{v^2}-2\right)+\frac{N}{2}\left(\frac{h^2}{v^2}\left(\frac{\phi}{f}-\frac{\phi_0}{f}\right)-\frac{\phi}{f}\right)\right.\\
&\left. +\frac{h^2}{v^2}\left(1+\left(\frac{\phi_0}{f}-\frac{\phi}{f}\right)\tan \frac{\phi_0}{f}-\frac{\cos \phi/f}{\cos \phi_0/f}\right)\right] +\frac{h}{v} \rho
\end{align}
where $N$ is a dimensionless parameter which is identified with Eq. (\ref{defM}) for the relaxion model.
We find that $\sin\theta= m_\phi/m_h$ as expected for models with a flat direction. The cubic parameter is given by
\be
U^{\prime \prime \prime}= -\frac{m^2_\phi}{f}\left(\tan \frac{\phi_0}{f} +\frac{6}{N}\right)
\ee
which must be positive for stable models.
This can be achieved, for instance,  by taking
\be
\phi_0= \frac{3\pi}{2} +\epsilon_0
\ee
where $\epsilon_0$ is positive and small. This guarantees than $m^2_\phi >0$ and $U^{\prime \prime \prime}>0$ as long as $\epsilon_0<6/M$.
We have represented the stabilisation of the flat direction by matter in the case of $\epsilon_0=10^{-2}$ and $M=100$ in Fig. (\ref{fig1}).
Screening depends on the variation of the mixing angle inside matter, which is given by
\be
\delta= -3\sin^2 \theta \epsilon
\ee
as a result the screening radius is given by
\be
R_{\rm scr}= \frac{1}{9\sin^4 \theta \epsilon^2 m_h }.
\ee
We can now briefly sketch the constraints imposed by laboratory and solar system experiments on the relaxion models.

\subsection{The allowed parameter space of nearly massless relaxion models}

\begin{figure*}
\begin{center}
\epsfxsize=9 cm \epsfysize=9 cm {\epsfbox{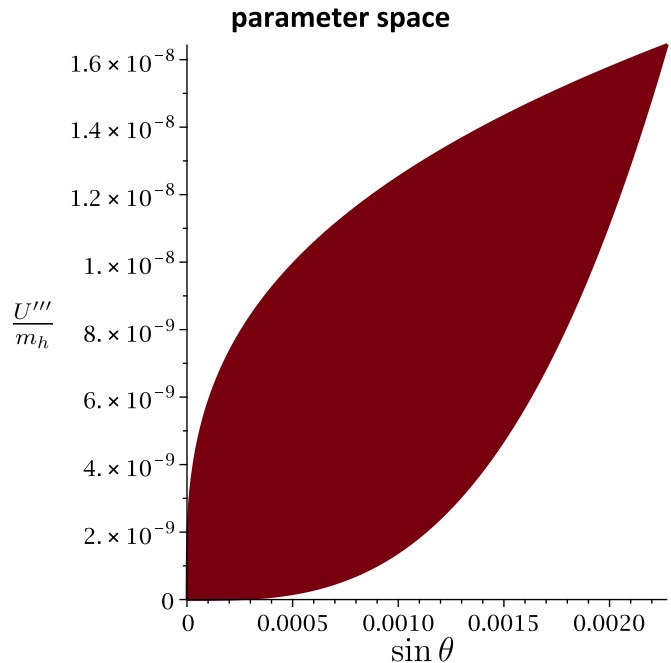}}
\end{center}
\caption{ The  parameter space of nearly massless relaxion potentials as a function of $\frac{U^{\prime \prime \prime}}{m_h}$ versus $\sin\theta$ that is stable under quantum corrections (the coloured region is allowed). The restrictions imposed by the quantum stability of the flat direction are stronger than the gravitational experimental bounds such as the Cassini test of fifth forces or the Microscope and Lunar Ranging  tests of the equivalence principle. }
\label{fig5}
\end{figure*}

We focus on models for which the relaxion is nearly massless, i.e. $\sin \theta_0\sim \frac{m_\phi}{m_h}$. This condition means that  the linear response theory fails and non-linear screening is potentially active. In this case the relaxion field is of nearly infinite range in vacuum.
The parameter space of nearly massless relaxion models is best described by two dimensionless parameters $\sin\theta$ and $U^{\prime \prime \prime}/m_h$.
A tight bound on $\frac{U^{\prime \prime \prime}}{m_h}$ is given by the condition on the quantum stability
of the flat direction which requires that
\be
\frac{U^{\prime \prime \prime}}{m_h} \lesssim (\sin \theta)^{1/3}\left(\frac{\rho}{m_h^3v}\right)^{1/3}.
\ee
The fermionic corrections to the flat direction are also under control provided that
\be
\frac{U^{\prime \prime \prime}}{m_h} \gtrsim \frac{m_\psi}{m_h}\sin^3 \theta.
\ee
The two quantum conditions are compatible when
\be
\sin \theta \lesssim \left(\frac{m_h}{m_\psi}\right)^{3/8} \left(\frac{\rho}{v m_h^3}\right)^{1/8}.
\ee
Taking as the mass of the fermion to be of order the top quark mass, $m_\psi= \sqrt 2 m_t$, and density of matter to be $\rho\simeq 1\ {\rm g/cm^3}$, we find that $\sin\theta \lesssim 10^{-3}$ for quantum corrections to be under control, see Fig. \ref{fig5} where the upper bound on $\sin\theta$ can be clearly seen.

Let us now discuss the screening radius. Objects of size $R$ are screened provided that
\be
R_{\rm scr} \lesssim R
\ee
For test objects in the laboratory, planets and the Sun, which have an average density of $\rho\simeq 10\ {\rm g/cm^3}$, one must impose
\be
\delta^2  \gtrsim \frac{1}{m_h R}
\label{con1}
\ee
which requires
\be
\frac{U^{\prime \prime \prime}}{m_h}\gtrsim  \frac{1}{\sin^5 \theta} \frac{v^4}{\rho}\frac{1}{vR}.
\label{cons1}
\ee
When combined with the quantum stability bounds, we find that no objects with sizes from those of  from the laboratory test masses up to stars of order the size of the  Sun are screened by the suppression due to a small effective charge $\beta_{\rm eff}$.
Hence the  non-linear screening is not strong enough to guarantee that laboratory tests of gravity, such as the ones of the  E\"otwash experiment or atomic interferometry and solar systems tests of gravity  are evaded. This can only be achieved if the mass of the scalar field in the near-vacuum of experiments with densities around $\rho_{\rm vac}\simeq 10^{-10} \ {\rm g/cm^3}$ is larger than typically $d\simeq 0.1 $ mm.
This imposes the constraint
\be
\frac{U^{\prime \prime \prime}}{m_h} \gtrsim \frac{v}{m_h} \frac{1}{\sin \theta} \frac{1}{\rho_{\rm vac} d^4}
\label{cons4}
\ee
Within the parameter space allowed by the quantum bounds, this is never restrictive. In fact, the Yukawa suppression is effective down to sizes around one nanometre.
Hence all laboratory tests are easily evaded.
As the mass of the  relaxion in matter scales with the density as $\rho^{1/4}$, we find that in the atmosphere with $\rho_{\rm atm}\simeq 10^{-4}\ \rm{g/cm^3}$ the range is smaller than 3 microns, in the galactic vacuum of density $\rho_{\rm atm}\simeq 10^{-23}\ \rm{g/cm^3}$ it is shorter than 20 centimeters,  and finally in the cosmological vacuum of density        $\rho_{\rm atm}\simeq 10^{-29}\ \rm{g/cm^3}$ the range is smaller than 10 metres.
In effect, although the relaxion is massless in vacuum, the presence of matter in the Universe even down to extremely low baryonic densities implies that the range of the relaxion is extremely short. This is enough to evade all the solar system tests.

The parameter space of nearly massless relaxion models can be seen in Fig. \ref{fig5}.  Restrictions on the parameter space of relaxion models come  from imposing the quantum stability of the flat direction. In Fig. \ref{fig5}, the mixing angle is constrained between $4\cdot 10^{-21}$ and $2\cdot 10^{-3}$ whilst $\frac{U^{\prime \prime \prime}}{m_h}$ can vary between $10^{-14}$ and $2\cdot 10^{-8}$. In all this parameter space, the range of the relaxion field is always short implying that all tests of gravity in the laboratory and the solar system are all always satisfied.

\section{Conclusions}
Light scalar fields coupled to matter through the Higgs portal, are a common component of theories that go beyond the standard models of particle physics and cosmology. As the light scalar mixes with the Higgs field, a light mode of the two field system can interact with matter, and mediate a long range fifth force. Such forces are tightly constrained by experiments and solar system observations and for light masses provide the tightest bounds on scalars coupled through the Higgs portal.

In this work we have shown that there are regions of parameter space for these models where the linearised treatment, usually used to analyse the phenomenology of these models, breaks down. This happens because the lightest mass eigenstate of the system becomes massless and it appears possible for both the Higgs and the light scalar to experience large field excursions. This also makes clear why it is not sufficient to `integrate out' the Higgs field and work with an effective theory for the light scalar; the Higgs field can be displaced from its vacuum expectation value, and those dynamics are important to capture the phenomenology of the theory.

Despite the apparent presence of a massless mode, the theory may remains perturbative and be stabilised at second order. As a result it remains possible to compute the strength of the long range fifth force mediated by the light mode. We find that in this non-linear regime the light mode couples less strongly to matter and so the fifth force is weaker and the field is less constrained by experiments for large enough bodies or large enough Newtonian potentials. This suppression of the fifth force by non-linearities in the theory is commonly known as screening.
Thus we have presented a novel way in which light scalar fields can exist in our universe, and couple to the fields of the Standard Model without being strongly constrained by experiment.

We have also commented on the implications of this non-linear regime for the relaxion model, which attempts to explain the Hierarchy Problem, and have identified the regime of parameter space in which non-linear effects need to be taken in to account.

\begin{figure*}
\epsfxsize=9 cm \epsfysize=9 cm {\epsfbox{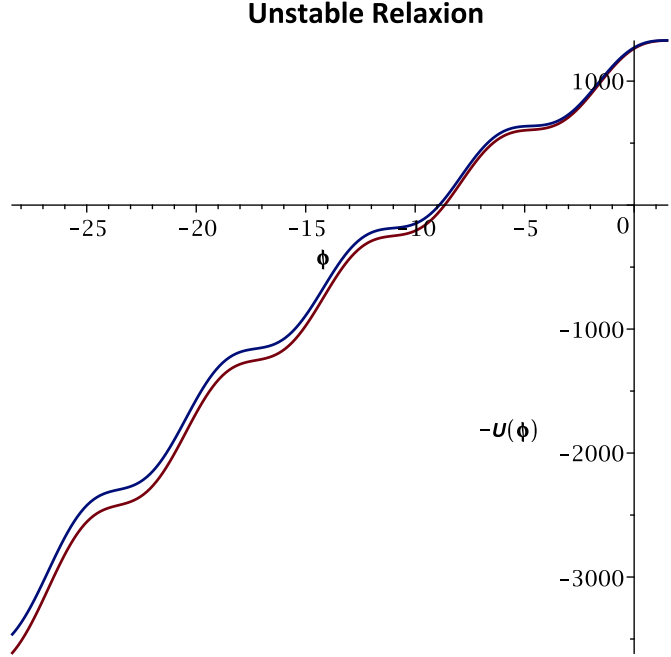}}
\caption{The unstable potential along the flat direction for a relaxion model defined by $\phi_0=(\frac{\pi}{2}-0.01)f$ and $N=100$. The potential with a flat direction (blue) is perturbed by matter (mauve). Static and spherical solutions around dense bodies are obtained by solving the spherical Klein-Gordon equation which corresponds to the motion, with friction, of a ball starting from the vacuum value at infinity. When no minima of $-U(\phi)$ exist at large and negative field values, no spherical solution can exist. Only time dependent solutions develop leading to a vacuum instability.  }
\label{fig2}
\end{figure*}

\section*{Acknowledgments}
We would like to thank the Light Scalars: Origin, Cosmology, Astrophysics and Experimental Probes meeting at the Centro de Ciencias de Benasque where the idea for this work was first sparked.  We would like to thank Brando Bellazzini for a careful reading of the manuscript and interesting questions, and Joerg Jaeckel for very helpful discussions.  CB is supported by a Research Leadership Award from the Leverhulme Trust and a Royal Society University Research
Fellowship.

\appendix

\section{Higgs-scalar models to second order}
\subsection{The minimum of the effective potential to second order}
\label{app:A}
We consider here a Higgs portal model with one scalar and a potential
\be
V_{\rm eff}(h,\phi)= V(h,\phi) +\frac{h}{v} \rho
\ee
in the presence of matter.
In vacuum, the vev's of the fields satisfy
\be
V_h(v,\phi_0)=0, V_{\phi}(v,\phi_0)=0
\ee
The perturbation by matter induces variations in the vev's which can be obtained to second order
\be
\delta h=- \frac{V_{\phi\phi}}{V_{\phi h}} \delta \phi - \frac{1} {V_{\phi h}} \left(V_{\phi\phi\phi}+ V_{\phi hh} \frac{V^2_{\phi\phi}}{V^2_{\phi h}} -2 V_{\phi\phi h} \frac{V_{\phi\phi}}{V_{\phi h}}\right)\frac{\delta\phi^2}{2}
\ee
leading to the equation for $\delta\phi$
\be
\lambda_- \delta \phi + U^{(3)} \frac{\delta\phi^2}{2} = \frac{V_{h\phi}}{V_{hh}} \frac{\rho}{v}
\label{min}
\ee
where
\be
\lambda_-=V_{\phi\phi} - \frac{V^2_{h\phi}}{V_{hh}}
\ee
and we have introduced the parameter
\be
U^{(3)} = V_{\phi\phi\phi} -\left(\frac{V_{h\phi}}{V_{hh}}+2\frac{V_{\phi\phi}}{V_{\phi h}}\right) V_{h\phi\phi} +3 \frac{V_{h\phi}}{V_{hh}}\frac{V_{\phi\phi}}{V_{h\phi}} V_{hh\phi} -  \frac{V_{h\phi}}{V_{hh}}\frac{V^2_{\phi\phi}}{V^2_{h\phi}}V_{hhh}.
\ee
All the derivatives are taken at $(v,\phi_0)$.
The minimum equation  of the main text  can be obtained from the effective potential
\be
U(\delta \phi)= \lambda_- \frac{\delta\phi^2}{2} + U^{(3)}\frac{\delta\phi^3}{6} - \frac{V_{h\phi}}{V_{hh}} \frac{\rho}{v} \delta\phi.
\ee
Linear perturbation theory holds when the quadratic term dominates. A non-linear solution exists when $U^{(3)}$ and $\frac{V_{h\phi}}{V_{hh}}$ are positive and the quadratic term is negligible.

\subsection{Properties of Higgs-Scalar mass matrices}
\label{app:B}
 The mass matrix of the models defined in appendix (\ref{app:A}) reads
\begin{equation}
\mathcal{M}^2 = \left(\begin{array}{cc}
V_{hh} & V_{h\phi}\\
V_{h\phi} & V_{\phi\phi}\\
\end{array}\right)
\end{equation}
where a derivative with respect to one of the fields is denoted by a subscript. We assume that $\vert V_{hh}\vert \gg \vert V_{\phi\phi}\vert $ and $\vert V_{hh}\vert \gg \vert V_{h\phi}\vert$. Within this approximation, the two eigenstates of this matrix are given by
\be
\lambda_+= V_{hh}, \ \ \lambda_-= V_{\phi\phi} - \frac{V^2_{h\phi}}{V_{hh}}.
\ee
Stability imposes that they should be both positive. The smallest eigenvalue is characterised by an eigenvector
\be
\vec \alpha_-= \left ( \begin{array} {c}
-\sin \theta\\
\cos\theta \\
\end{array}
\right )
\label{eiga}
\ee
where the mixing angle is given by
\be
\tan \theta= \left(1- \frac{\lambda_-}{V_{hh}}\right)^{-1}\frac{V_{h\phi}}{V_{hh}}
\ee
In the large $V_{hh}$ limit this is simply
\be
\tan \theta\simeq \sin \theta \simeq \frac{V_{h\phi}}{V_{hh}}.
\ee
The mass matrix can be diagonalised as
\be
{\cal M}^2= R^T_\theta {\cal M}_D^2 R_\theta
\ee
where
\begin{equation}
\mathcal{M}^2_D = \left(\begin{array}{cc}
\lambda_+ & 0\\
0 & \lambda_-\\
\end{array}\right)
\end{equation}
and the rotation matrix is
\begin{equation}
R_\theta = \left(\begin{array}{cc}
\cos\theta & \sin\theta\\
-\sin\theta & \cos\theta\\
\end{array}\right)
\end{equation}
When the vacuum of the theory, obtained as a minimum of the effective potential in the absence of matter, is perturbed by the small matter contribution, see (\ref{min}), the eigenvalues and eigenvectors of the mass matrix are perturbed.
The perturbation to the lowest eigenmass is
\be
\delta \lambda_-= \lambda_\phi^{(1)} \delta\phi
\ee
where we have defined
\be
\lambda_\phi^{(1)} = V_{\phi\phi\phi} -\left(2\frac{V_{h \phi}}{V_{hh}}+\frac{V_{\phi\phi}}{V_{h\phi}}\right) V_{h\phi\phi} +\left( 2\frac{V^2_{h \phi}}{V^2_{hh}}+\frac{V_{\phi \phi}}{V_{hh}}\right)  V_{hh\phi} -  \frac{V_{\phi \phi}}{V_{hh}}\frac{V_{h \phi}}{V_{hh}}V_{hhh},
\ee
The variation of the mixing angle is given by
\be
\delta \theta= \left(V_{h\phi\phi} - V_{hh\phi}\left( \frac{V_{h\phi}}{V_{hh}}+\frac{V_{\phi\phi}}{V_{h\phi}}\right) + V_{hhh} \frac{V_{\phi\phi}}{V_{hh}}\right) \frac{\delta \phi}{V_{hh}}.
\ee
We apply these results to the case where the potential is given by $V(h,\phi)=- \frac{\mu^2(\phi)}{2} h^2 +\lambda \frac{h^4}{4}$ in the main text.
\subsection{The effective potential in the nearly massless case}
When $\lambda_-$ is nearly vanishing, i.e. when the quadratic term in $U(\delta\phi)$ is subdominant then we have
\be
V_{\phi\phi}V_{hh}\simeq V_{h\phi}^2.
\ee
In this case we have also the identity
\be
U^{(3)}\simeq \lambda_\phi^{(1)} \simeq U^{\prime\prime \prime}
\ee
where
\be
U^{\prime\prime \prime}=V_{\phi\phi\phi} -3\sin\theta V_{h\phi\phi} +3 \sin^2\theta\frac{V_{\phi\phi}}{V_{h\phi}} V_{hh\phi} -  \sin^3 \theta V_{hhh}
\ee
is the third derivative of the potential $V(h,\phi)$ along the direction $\vec \alpha_-$, i.e. we have
\be
U(\delta\phi)= V(v-\sin\theta \delta\phi, \phi_0 +\delta \phi)-V(v,\phi_0)
\ee
from which we can identify the perturbation to the lowest mass eigenvalue
\be
\delta \alpha_-= \frac{d^2U}{d\delta^2 \phi}= U^{\prime\prime \prime}\delta\phi.
\ee
Stability imposes to choose as perturbed minimum the one with $\delta\phi>0$.

\section{The coefficient $B_-$}
\label{app:C}

The expression for $B_-$ in the general case is given by
\be
\frac{B_-}{\sqrt y_- R} = b - \frac {A}{\det X}
\ee
where
\be
\det X= c^2_\delta \left(c_++z_+ \frac{s_+}{x_+}\right)\left( c_-+z_- \frac{s_-}{x_-}\right)+s_\delta^2 \left(c_++z_- \frac{s_+}{x_+}\right)\left(c_-+z_+ \frac{s_-}{x_-}\right)
\ee
and
\be
A= (1+z_+) \left(c_+ \frac{s_-}{x_-} -c_- \frac{s_+}{x_+}\right) c_\delta s_\delta a + (1+z_-) \left(z_+ \frac{s_+}{x_+} \frac{s_-}{x_-}+ s_\delta^2 c_- \frac{s_+}{x_+}+ c_\delta^2 c_+ \frac{s_-}{x_-}\right) b.
\ee
We have defined
\be
a= \cos\theta_o \delta h+ \sin\theta_o \delta \phi, \ \ b= -\sin\theta_o \delta h+ \cos\theta_o \delta \phi
\ee
and $s_\pm= \sinh x_\pm$, $ c_\pm= \cosh x_\pm$, $c_\delta= \cos \delta$, $s_\delta= \sin \delta$ where $x_\pm = \sqrt{\lambda_\pm} R$. We have also defined
$z_\pm =\sqrt{y_\pm} R$.
In the limit $x_+\gg 1$ and $ x_- \ll 1 $ with $z_+\sim x_+$ we find that
\be
\frac{B_-}{{\sqrt y_-} R}= \frac{x_-^2}{3} \frac{b}{ c_\delta^2 + s_\delta^2 \frac{z_+}{2}} - \frac{ y_+ s_\delta } { 2 c_\delta^2 + s_\delta^2 {z_+}}(  \cos \theta_i \delta h + \sin \theta_i \delta \phi)
\ee
where $\theta_i= \theta_o -\delta$.
Expanding
\be
\cos \theta_i \delta h + \sin \theta_i \delta \phi\simeq - \sin \delta \delta \phi
\ee
at linear order in $\delta$.
Combining these results we find that
\be
\frac{B_-}{{\sqrt y_-} }= \frac{\frac{x_-^2}{3}+ s_\delta^2 \frac{x_+}{2}}{c_\delta^2 + s_\delta^2 \frac{x_+}{2}}R \delta \phi
\ee
We use this expression in the main text to discuss the screening of the Higgs portal.

\section{The unstable vacuum}
\label{app:D}

\begin{figure*}
\epsfxsize=9 cm \epsfysize=9 cm {\epsfbox{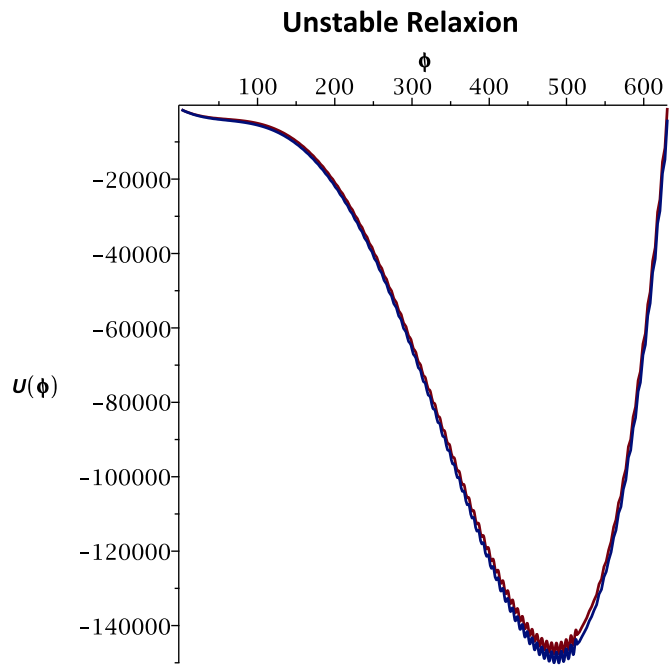}}
\caption{The unstable potential along the flat direction for a relaxion model defined by  $\phi_0=(\frac{\pi}{2}-0.01)f$ and $N=100$. The potential with a flat direction (mauve) is perturbed by matter (blue) and develops minima far away in field space compared to the vacuum value. Cosmologically, the field will be destabilised by the matter component of the Universe and settle in a stable minimum of the potential. }
\label{fig3}
\end{figure*}

We are interested in the situation where the Higgs mass dominates, i.e. $m_h\gg m_\phi$, and the potential admits a flat direction starting at $(h,\phi)=(v,\phi_0)$ along the massless direction parameterised by the eigenvector $\vec\alpha_-$. When the cubic parameter $U^{\prime\prime\prime}$ is negative, the presence of matter detabilises the flat direction, see Fig.(\ref{fig3}).  This destabilisation was previously noticed in Ref.~\cite{Budnik:2020nwz}. The dynamics of the Higgs-scalar system along this valley can be simplified by decoupling the fast mode along the eigenvector $\vec\alpha_+$ of the largest mass eigenvalue $\alpha_+$ and the slow mode along the vector $\vec\alpha_-$. As the fields move away from the vacuum value, the slow mode lie at the bottom of the valley constructed as the integral curve tangent to $\vec\alpha_-$.   The valley starting at $(h,\phi)=(v,\phi_0)$  is parameterised as
$
\frac{d}{ds} \vec v = \vec \lambda_-(\vec v)
$
where
the field values along this valley are given by
\be
\vec v (s)=\left ( \begin{array} {c}
h(s)-v\\
\phi(s)-\phi_0\\
\end{array}
\right )
.
\ee
It is more convenient to parameterise the valley  by $\phi$ after eliminating the dummy parameter $s$. Along the valley the potential for the slow mode can be expressed as
\be
U(\phi-\phi_0)= V(h(\phi),\phi)-V(v,\phi_0) + \frac{h(\phi)-v}{v} \rho
\ee
in the presence of matter. When approximating the valley as a straight line as $m_h\gg m_\phi$, we retrieve that $h(\phi)-v= -\sin\theta (\phi-\phi_0)$ and the potential
$U(\delta\phi)$ that we expanded to cubic order in the main text.
Spherical solutions correspond to the motion of a ball in a potential $-U(\phi)$ with friction. If we assume that $-U(\phi)$ does not admit stable minima for $\phi<\phi_0$, no spherical and static solution can interpolate between the vacuum value $\phi_0$ at infinity and another minimum of $-U(\phi)$ in the centre of the overdensity. This is the case of relaxion models as can be seen in Fig. (\ref{fig2}). On the contrary, the presence of the matter density $\rho$ destabilises the vacuum and a time dependent solution develops. As there is no friction, the field will carry out large oscillations, see Fig. (\ref{fig3}) for the relaxion case. As the potential is  flat around the vacuum value, the field  lingers around the origin for a time given by
\be
t_{\rm ins}= \frac{4\sqrt 3}{\sqrt \vert U^{\prime\prime\prime}\vert} \left(\frac{v}{6\rho}\right)^{1/4}.
\ee
For a body of density $10\ {\rm g/cm^3}$, this is around $10^{-20}$ seconds for $U^{\prime\prime\prime}=1 $ GeV. Hence for non-tuned values, the field starts oscillating and a bubble of oscillating field forms around the body. It is likely that this bubble eventually expands and fills all the Universe. In fact the vacuum configuration would certainly have been destabilised by the matter density of the Universe in the matter era and after damped oscillations due to the Hubble friction the field would have settled to one  stable vacuum, see Fig.(\ref{fig3}) for the relaxion case. In conclusion, models with an unstable  flat direction are not physical.

\bibliographystyle{utphys}
\bibliography{relaxion_screening_arXiv}

\end{document}